\documentclass[12pt]{article}
\title{
Analytic Methods in Nonperturbative QCD}
\author{Yu.A.Simonov\\ State Research Center\\Institute of
Theoretical and Experimental Physics, \\ Moscow, Russia}
 \date{}

\newcommand{\be}{\begin{equation}}
\newcommand{\ee}{\end{equation}}

\def\fun#1#2{\lower3.6pt\vbox{\baselineskip0pt\lineskip.9pt
\ialign{$\mathsurround=0pt#1\hfil
##\hfil$\crcr#2\crcr\sim\crcr}}}

\newcommand{\ver}{\mbox{\boldmath${\rm r}$}}
\newcommand{\vesig}{\mbox{\boldmath${\rm \sigma}$}}
\newcommand{\vep}{\mbox{\boldmath${\rm p}$}}

\newcommand{\vexi}{\mbox{\boldmath${\rm \xi}$}}
\newcommand{\veta}{\mbox{\boldmath${\rm \eta}$}}

\newcommand{\lan}{\langle}
\newcommand{\ran}{\rangle}
\newcommand{\lll}{\langle}
\newcommand{\rrr}{\rangle}

\newcommand{\T}{\mbox{Tr}}
\begin{document}

\maketitle

\begin{abstract}

Recently developed analytic methods in the framework of the Field
Correlator Method are reviewed in this series of four lectures and
results of calculations are compared to lattice data and experiment.
Recent lattice data demonstrating the Casimir scaling of static quark
interaction strongly support the FCM and leave very little space for
all other theoretical models, e.g. instatnton gas/liquid model.
Results of calculations for mesons, baryons, quark-gluon plasma and
phase transition temperature demonstrate that new analytic  methods
are a powerful tool of nonperturbative QCD along with lattice
simulations.

\end{abstract}

\section{General QCD formalism for Green's functions.}

1. Starting to work in QCD one should have in mind some specific
features, which make this theory unique; two of these are most
important.

In QCD (as in any local gauge theory in the confining phase) it is
impossible to introduce asymptotic states of quarks and gluons, hence
the popular formalism of LSZ \cite{1} is not applicable. Therefore
one works with off-shell quantities, like the Green's functions and
later on one introduces Hamiltonian and scattering amplitudes.

2. In QCD the Nonperturbative (NP) fields are Euclidean (c.f.
\cite{2})
and therefore one should work out the formalism of Euclidean vacuum
averaging first, and later one goes over to the Minkowskian space
(inevitable for scattering amplitudes.).

In this section we introduce first simple general and exact formulas
for Green's functions, Hamiltonians etc. and later discuss possible
approximations.

Consider a general partition function of QCD
\be
Z(J_\mu, \eta, \bar \eta)= \int DA_\mu D\psi D\psi^+
e^{-S(A,\psi,\psi^+) + \int(J_\mu A_\mu+ \eta^+\psi+\psi^+\eta)d^4x}
\label{1}
\ee
where $S$ is the standard QCD Euclidean action.
   \be
   S=\frac12 \int tr F^2_{\mu\nu} d^4x-\int \psi^+(i\hat \partial +im
   +g\hat A)d^4x.
   \label{2}
   \ee
   Here gauge fixing terms and Faddeev-Popov ghosts are suppressed
   for simplicity.

   A spicific feature of local gauge theories is that only
   gauge-invariants can enter as physical states, and local gauge
   invariance cannot be spontaneously violated (the Elitzur theorem).
   Therefore the only way to proceed is to form gauge-invariant
   Green's functions from (\ref{1}). E.g. for the $\bar q q$ state
   one might have a local or nonlocal meson Green's function.
   \be
   G^{(M)}_{f,in} (xx|yy)=\lan \psi^+(x)
   \Gamma_f\psi(x)\psi^+(y)\Gamma_{in}\psi(y)\ran,~~ {\rm or}\lan
   \psi^+\Gamma_f\Phi\psi \psi^+\Gamma_{in} \Phi \psi\ran
   \label{3}
   \ee
   where $\Gamma_f,\Gamma_{in}=1,\gamma_5, \gamma_\mu,...; \Phi(x,y)
   = P\exp ig \int^x_yA_\mu dz_\mu,$ and $\lan ... \ran$ means
   averaging over $DA_\mu D\psi D\psi^+$ as in (\ref{1}).

   Integrating over $D\psi D\psi^+$ in (\ref{1}) one obtains
   for the meson Green's function.
   $$
   G^{(M)} = \int DA_\mu e^{-S_0(A)} \det (m+\hat
   D)\{tr_{\alpha}[\Gamma_f S(x,y) \Gamma_{in} S (y,x)]- $$
 \be   -tr_{\alpha} (S(x,x) \Gamma_f)tr_\alpha (S(y,y)
 \Gamma_{in})\},
  \label{4}
   \ee
    with notatins   for the quark Green's
     function
     \be
     S(x,y) =\lan x|(m+\hat D)^{-1}|y\ran.  \label{5}
     \ee
     In QED one can define for $G^{(M)}$ the Bethe-Salpeter (BS) or
     similar equations to take into account the proper subclass
                 of perturbative diagrams. In QCD this way is
                 impossible since both BS and Dyson-Schwinger eqs are
                 not gauge-invariant, whenever one approximates the
                 kernels by finite sets of diagrams.

                 For gluons one should introduce gauge-invariant
                 initial and final states made of $F_{\mu\nu},
                 D_\mu$.
E.g. for local states $\Psi_{f(in)} = tr F_{\mu\nu} (x)
F_{\alpha\beta} (x)$  or $tr D_\rho F_{\mu\nu} (x) F_{\alpha\beta}
(x))$ etc. and define Green's functions
\be
G^{(Gl)}_{f,in}=\lan \Psi^+_f(x) \Psi_{in} (y)\ran.
\label{6}
\ee
Integration over $D\psi D\psi^+$ allows to obtain only the form
\be
G^{(Gl)}_{f,in}=\int DA_\mu e^{-S_0(A)}\det (m+\hat D) \Psi^+_f(x) \Psi_{in}
 (y).
   \label{7}
   \ee
   To proceed one should use the Background Field Formalism (BFF) and
   the Feynman-Schwinger (or world-line) Representation (FSR) which
   will be discussed in the next lectures.

   Here we shall give connection to the Hamiltonian technique, which
   is useful both for bound states and for scattering.

   The following general relation holds true for any Green's function
   \be
   G^{(k)}_{f,in} (x,y|x', y')=
    \lan f, x,y|e^{-HT} | in, x', y'\ran=
    \lan f, x,y|n>e^{-M_nT}<n | in, x', y'\ran
    \label{8}
    \ee
    with $M_n,|n>$ being eigenvalues and eigenfunctions of $H$
     respectively.
    Here we have defined a set of hypersurfaces $\{\sum_t\}$ on which
   the Hamiltonian is constructed, and the evolution parameter $t$,
   perpendicular  to $\sum_t$, which varies between, say, $t=0$ for
   the in-state and $t=T$ for the final state.

   The Hamiltonian $H$ is the general QCD Hamiltonian which is known
   in several gauges, e.g. Coulomb \cite{3}, temporal \cite{4} or in
   polar representation \cite{5}. It is a matrix, containing creation
   and destruction of quarks and gluons, and angular brackets
   in (\ref{8}) do single out the initial and final indices of this
   matrix.

   The most important feature of QCD is confinement, which means that
   some part of gluons are condensed in the form  of strings between
   charges. Unfortunately the general QCD Hamiltonians
   \cite{3}-\cite{5} are not yet helpful in understanding confinement
   properties of QCD, and one cannot use it as a tool in exploring
   vacuum properties.

   Therefore our main aim would be to formulate effective Hamiltonian
   for the valence (noncondensed) gluons and quarks.

   The way to do it will be described in next lectures.

   \section{Background Field Formalism}

 One of basic mysteries of QCD is the fact, that gluon field plays
 two different roles:

i) gluons are propagating, and at small distances this process can
 be described perturbatively, leading in particular  to color Coulomb
 interaction between quarks (antiquarks);

 ii) gluons form a kind of condensate, which serves as a background
 for the propagating perturbative gluons and quarks. This background
 is Euclidean and ensures  phenomena  of  confinement and CSB.

 Correspondingly we shall separate  the total gluonic field $A_\mu$
 into perturbative part $a_\mu$ and nonperturbative (NP) background
 $B_\mu$:
 \be
 A_\mu=B_\mu+a_\mu
 \label{9}
 \ee

 There are many questions about this separation, which may be
 answered now only partially, e.g. what exactly is the criterion of
 separation. Possible answer is that perturbative fields $a_\mu$ get
 their dimension from distance (momentum), and   therefore all
 correlators of   fields $a_\mu$ (in absence of $B_\mu$) are singular
 and made of
 inverse powers of ($x-y)$ and logarithms, where enters the only
 dimensional parameter of perturbative QCD -- $\Lambda_{QCD}$.
 Therefore evidently any dimensionful constant, like hadronic masses
 or string tension cannot be obtained as a perturbation series. In
 contrast to that, NP fields  $B_\mu$ have mass dimension due to the
 violation of scale invariance
 which is  intrinsically  present in the
 gluodynamics Lagrangian. The origin of separation (\ref{9})) is
 clearly seen in the  solutions of nonlinear equations for field
 correlators \cite{6}:  a perturbative  solution of those leads  to
 singular power-like field correlator, whereas at large distances
 there is a selfconsistent solution  of the equations, decaying
 exponentially with distance with arbitrary mass scale, since
 equations in  \cite{6} are scale--invariant. Full solution
 including intermediate distances produces mixed
 perturbative--nonperturbative terms,  containing both inverse powers
 of distance and exponentials.  For these terms criterion of
 separation fails.

 One can avoid formally the question of separation principle (and of
 double counting) using t'Hooft identity \cite{7}, which allows to
 integrate in (1) independently over $B_\mu$ and $a_\mu$:
 \be
 Z=\frac{1}{N'}\int DB_\mu \eta (B) D\psi D\bar \psi  D a_\mu
 e^{L(B+a)}
 \label{10}
 \ee
 Here $L(A)=L_0+L_1+L_{int}$, and  $A_\mu$ is taken to be
 $B_\mu+a_\mu$;
 the  weight $\eta (B)$ is arbitrary and may be  taken as constant.
  For the exact formalism starting from (\ref{10}) we
 refer the reader to \cite{7,8}, and here we only quote the form of
 background propagator $G_{\mu\nu}$ of the gluon, which is found from
 $L(B+a)$ and will be used in what follows,
  \be
 G_{\mu\nu}=(D^2_\lambda \delta_{\mu\nu} +2ig F_{\mu\nu})^{-1}
 \label{11}
 \ee
 where $D_\lambda=\partial_\lambda-ig B_\lambda$, and $F_{\mu\nu} =
 \partial_\mu B_\nu-\partial_\nu B_\mu -ig [B_\mu, B_\nu]
 $.

To define the perturbation theory  series in $ga_\mu$ one starts from
(\ref{10}) and  rewrites the Lagrangian as follows:
$$
L_{tot}=L_{gf}+L_{gh}+L(B+a)=
$$
\be
L_0+L_1+L_2+L_{int}+L_{gf}+L_{gh}
\label{12}
\ee
where $L_i$ have the form:
\begin{eqnarray}
L_2(a)&=&\frac{1}{2} a_{\nu}(\hat{D}^2_{\lambda}\delta_{\mu\nu} -
\hat{D}_{\mu}\hat{D}_{\nu} + ig \hat{F}_{\mu\nu}) a_{\mu}= \nonumber \\
&=&\frac{1}{2} a^c_{\nu}[D_{\lambda}^{ca}D_{\lambda}^{ad} \delta_{\mu\nu}
- D_{\mu}^{ca}D_{\nu}^{ad} - g~f^{cad}F^a_{\mu\nu}]a^d_{\mu}~~,
\label{13}
\end{eqnarray}
\begin{eqnarray}
D_{\lambda}^{ca} &=&
 \partial_{\lambda}\cdot \delta_{ca}+ g~f^{cba} B^b_{\lambda}
\equiv \hat{D}_{\lambda}
\nonumber
\\
L_0 &=& -\frac{1}{4}
 (F^a_{\mu \nu}(B))^2 ~;~~~ L_1=a^c_{\nu} D_{\mu}^{ca}(B) F^a_{\mu\nu}
\nonumber
\\
L_{int} &=& -\frac{1}{2} (D_{\mu}(B)a_{\nu} -D_{\nu}(B)a_{\mu})^a
g~f^{abc} a_{\mu}^b a_{\nu}^c - \frac{1}{4} g^2 f^{abc} a_{\mu}^b a_{\nu}^c
f^{aef} a^e_{\mu} a^f_{\nu}
\label{14}
\end{eqnarray}
\begin{eqnarray}
L_{ghost}=-\theta^+_a (D_{\mu}(B)D_{\mu} (B+a))_{ab} \theta_b.
\label{15}
\end{eqnarray}
To calculate Green's functions of any hadrons one can use
(\ref{10}) and write
 $$ G_h(X;Y)= const \int DB_\mu\eta(B)D\psi
D\bar \psi Da_\mu \Psi^+_f(X)\Psi_{in} (Y) e^{L_{tot}} $$ \be \equiv
\langle \Psi^+_f(X)\Psi_{in} (Y)\rangle _{B,\psi,a}.
\label{16}
 \ee
  Here
$\Psi_{f, in} $ are initial and final hadron states made of
$B,\psi,\bar\psi,a$. To calculate
 the  integral $D{a_\mu}$ one can use
perturbative expansion in $ga_\mu$, i.e. neglect in first
approximation $L_{int}$, containing terms  $a^3, a^4$ and take into
account only quadratic terms $L_2, L_{gf}, L_{gh}$.
The  contribution of
$L_1$ was studied in \cite{8} and it was shown there that for the most
processes (e.g. for hadron Green's functions) it can be neglected
(with accuracy of about better than 10\%).

It is convenient to choose background gauge for $a_\mu$, $D_\mu
a_\mu=0$, and take gauge transformation in the form

\be
B_\mu\to U^+(B_\mu+ \frac{i}{g} \partial_\mu)U, a_\mu\to U^+ a_\mu U
\label{17}
\ee
Now one can choose $\Psi_f, \Psi_{in}$ as gauge-invariant
 forms built from $B_\mu, a_\mu,\psi,\bar \psi$ and
 to average over $B_\mu,\psi,\bar \psi, a_\mu$ as shown in
 (\ref{16}).  Integrating over $DB_\mu$ might seem an impossible
adventure, but we shall see that it can be
always written as products of  Wilson loops with some insertions,
which can be treated in two ways:

i) or in  the form of the area law. Then contribution of fields
$B_\mu$ is  given  and fixed by the string tension $\sigma(B)$.
Equating $\sigma(B)= \sigma_{exper}$ one fixes contribution of the
background fields.

ii) Using cluster expansion, and expressing result through lowest
correlators, e.g. $\langle FF\rangle $, or
its scalar parts $D(x)$ and $D_1(x)$. In this case
background is fixed by these functions taken as an input, e.g. from
lattice data.

The rest integrals, over $Da_{\mu} D\psi D\bar \psi$ are Gaussian and
can be
simply done. Now we turn to the construction of $\Psi_{f, in}$.

One can choose local or nonlocal forms for $\Psi_{in,f}$, the latter
are obtained by insertion of $\Phi(x,y)$  in the local
expressions, which are

$\Psi_{in,f}=\bar \psi \Gamma_i \psi$ for mesons, $\Gamma_i=1,
\gamma_5, \gamma_\mu, \gamma_\mu\gamma_5, \sigma_{\mu\nu}$

$\Psi_{in,f}=\bar \psi \Gamma_i f (a, Da)\psi $ for hybrids,
where $f$ is a polynomial, the simplest form is
$\bar \psi \gamma_i a_\mu\psi$

$\Psi_{in,f}=tr a\Lambda_2 a$ and $tr(\Lambda_3 a^3)$ for
glueballs made of 2 and 3 gluons  respectively. $\Lambda_2,
\Lambda_3$ are polynomials made of $D_\mu(B)$ and ensure the
needed tensor structure for given quantum numbers. In the
simplest case $\Lambda_2=\Lambda_3=1$.

$\Psi_{in,f}=e_{abc} \psi^a_{f_1\alpha} \psi^b_{f_2\beta}
 \psi^c_{f_3\gamma} K^{
 f_1\alpha,f_2\beta,f_3\gamma}$ for baryons, where $a,f_i,\alpha$ are
 color, flavour and Lorentz index respectively.

 It is clear that higher hybrid states for mesons and baryons are
 obtained by adding  factors of $a_\mu$ in $\Psi_{in, f}$.

 Insertion of given $\Psi_{in, f}$ in (\ref{16}) yields after
 integration over $Da_\mu D\psi D\bar\psi$ the hadron Green's
 function, which can be written  respectively  for mesons $$
 G_M(X,Y)=\langle tr(\Gamma_i^{(f)}(X)G_q(X,Y)
 \Gamma_k^{(in)}(Y)G_{\bar q}(X,Y))\rangle _B
 $$
   \be
 -\langle tr(\Gamma_i^{(f)}(X)G_q(X,X))
 tr (\Gamma_k^{(in)}(Y)G_{\bar q}(Y,Y))\rangle _B,
   \label{18}
   \ee
   where $G_{q,\bar q}$ is the quark (antiquark) Green's function,
   \be
   G_q(X,Y)=(\hat \partial+m-ig\hat B)^{-1}_{X,Y},
   \label{19}
   \ee
   for hybrids
   \be
 G_{hyb}(X,Y)=\langle tr(\Gamma_i^{(f)}(X)G_q(X,Y)
 G_g(X,Y)\Gamma_k^{(in)}G_{\bar q}(X,Y))\rangle _B,
 \label{20}
 \ee
   where $G_g$ is the gluon Green's function (\ref{11}).

   For glueballs one has
   \be
 G_{glueball}=\langle tr(\Lambda_2^{(f)}(X)G_g(X,Y)
 \Lambda_2^{(in)}(Y)G_g(X,Y))\rangle _B
 \label{21}\ee
   and for baryons
   \be
 G_B(X,Y)=\langle tr(eK^{(f)}\prod^3_{i=1}G_q^{(i)}(X,Y)  eK^{(in)}
 )\rangle _B
 \label{22}
 \ee

 The structure of hadron Green's function is clear: it is a product
 of quark and gluon Green's function averaged over the background
 field $B_\mu$. In the next section we shall discuss the way, how to
 make this averaging explicit and express it through the Wilson loop.

      \section{Feynman-Schwinger representation}

   The present stage of development of field theory in general and of
   QCD in particular requires the exploiting of nonperturbative
   methods in addition to summing up perturbative series. This calls
   for specific methods where dependence on vacuum fields can be made
   simple and explicit. A good example is provided by the so-called
   Feynman-Schwinger Representation (FSR) \cite{9,10}
   based on the
   Schwinger proper time  and Feynman path integral
   formalism \cite{11,12}.

   For  some earlier references see \cite{10} and \cite{13}.
   More
   recently some modification of FSR was exploited in \cite{14} and
   \cite{15}.  The
   double-logarithm  perturbative summation of  amplitudes is
   especially convenient for FSR and was extensively studied in
   \cite{16}.

   Meanwhile the first extension of FSR to nonzero temperature field
   theory was done in \cite{17}. That
   was the basis for the systematic study of the role of
   nonperturbative (NP) configurations in the temperature phase
   transitions \cite{18,19,20,21}.

   One of the most important advantages of FSR is that it allows to
   reduce  physical amplitudes to the weighted integrals of the
   averaged Wilson loops. Thus the field (both perturbative and NP)
   enter only through Wilson loops and for the latter one can apply
   cluster expansion theorem \cite{22} which allows to sum up a
   series of approximations directly in the exponent.

   Therefore one can avoid Feynman diagram summation to get
   asymptotics of formfactors, as it was done in \cite{16}.

   The role of FSR in the treatment of NP effects is more crucial. In
   this case one  can develop a powerful method of Background
   Perturbation Theory  treating NP fields as a background
   \cite{7,8}.

   We start with the scalar particle (e.g. Higgs boson) interacting
   with the nonabelian vector potential via Euclidean Lagrangian
   \be
   L_\varphi = |D_\mu \varphi|^2 \equiv |(\partial_\mu-ig
   A_\mu) \varphi |^2
   \label{1.a}
   \ee
   so that the Green's function of $\varphi$ can be written using the
   Fock--Schwinger proper time representation as
   \be
   G(x,y) = (m^2-D^2_\mu)^{-1}_{xy} = \langle x| P \int^\infty_0 ds
   e^{-s(m-D^2_\mu)}|y\rangle.
   \label{2.a}
   \ee
   The next step can be done as in \cite{9}, treating $s$ as the
   ordering parameter. Note the difference of the  integral
   (\ref{2.a}) from the case of QED treated in
   \cite{10,12,13}:  $A_\mu$ in our case is the
   matrix operator $A_\mu(x)= A_\mu^a(x)T^a$ and it does not commute
   for different $x$, hence the ordering operator $P$ in (\ref{2.a}).
   The exact meaning of $P$ becomes clear in the final form of the
   path integral
    \be G(x,y) =\int^\infty_0 ds (Dz)_{xy} e^{-K} P \exp
   (ig \int^x_y A_\mu (z) dz_\mu)
    \label{3.a}
     \ee
     where $K=\frac14
    \int^s_0 d\tau (\frac{dz_\mu}{d\tau} )^2+m^2s$, and
    \be (Dz)_{xy}
   = \lim_{N\to \infty} \prod^N_{n=1}\int \frac{d^4
   z(n)}{(4\pi\varepsilon)^2} \int \frac{d^4p}{(2\pi)^4} e^{ip
   (\sum^N_{n=1} z(n)-(x-y))}, ~~N\varepsilon =s.  \label{4.a} \ee

   The last integral in (\ref{4.a}) ensures that the path
   $z_\mu(\tau), 0\leq \tau\leq s$, starts at $z_\mu(0) = y_\mu$ and
   ends at $z_\mu(s) =
   x_\mu$ The form of (\ref{3.a}) is the same as in
   the case of QED except for the ordering operator $P$ which gives
   exact meaning to the integral of the noncommuting matrices
   $A_{\mu_1}(z_1), A_{\mu_2}(z_2)$ etc. In the case of QCD the form
   (\ref{3.a}), (\ref{4.a}) was introduced in \cite{9}.

   The usefulness of the FSR (\ref{3.a}) becomes clear when one
   considers the physical amplitude, e.g. the Green's function of the
   white state $tr(\varphi^+(x) \varphi(x))$ or its nonlocal version
   $tr[\varphi^+ (x) \Phi(x,y) \varphi(y)]$, where $\Phi(x,y)$ to be
   widely used in what follows is parallel transporter along some
   arbitrary contour $C(x,y)$
   \be
   \Phi(x,y) =P\exp (ig \int^x_y A_\mu(z) dz_\mu).
   \label{5.a}
   \ee

   One has by standard rules
   $$
   G_\varphi (x,y) =\langle
    tr (\varphi^+(x) \varphi(x))
    tr (\varphi^+(y) \varphi(y))\rangle_A=
     $$
     \be
     =\int^\infty_0 ds_1
     \int^\infty_0 ds_2
   (Dz)_{xy}
   (Dz')_{xy}
               e^{-K-K'} \langle W\rangle_A+...
   \label{6.a}
   \ee
   where dots stand for the disconnected part,
   $\langle G_\varphi (x,x) G_\varphi (y,y)\rangle_A$ and
    we have used the fact that the propagator of the
   charge-conjugated particle $\varphi^+$ is proportional to
   \be
   \Phi_C(x,y) =P\exp (ig \int^x_y A_\mu^{(C)} (z) dz_\mu), ~~
   A_\mu^{(C)}=-A_\mu^{(T)}
   \label{7.a}
   \ee
   and therefore the ordering $P$ must be inverted, $\Phi_C(x,y)
   =P^{(T)} \exp (ig \int^y_x A_\mu (z) dz_\mu)$.

    Thus all dependence on $ A_\mu(x)$ in $G_\varphi$ is reduced to
    the Wilson loop average
     \be \langle W\rangle_A=\langle tr P_C \exp
     ig \int_C A_\mu (z) dz_\mu \rangle_A .
     \label{8.a}
      \ee Here $P_C$
      is the ordering  around the closed loop $C$ passing through the
      points $x,y$; the loop being made of the path $z_\mu(\tau)$ and
      $z'_\mu(\tau')$ to be integrated
      over in the integral (\ref{6.a}).

      Consider now a  fermion (quark) Green's function in the
      Euclidean external gluonic field. Similarly to   the QED case
      considered in \cite{10} one has
      $$G_q(x,y)= \langle \psi(x) \bar \psi(y)\rangle_q=\langle x
      |(m_q+\hat D)^{-1}|y\rangle =
      $$
      $$
      =\langle x|(m_q-\hat D)(m^2_q-\hat D^2)^{-1}|y\rangle=
      $$
      \be
      (m_q-\hat D)\int^\infty_0 ds (Dz)_{xy} e^{-K} \Phi_\sigma(x,y)
      \label{9.a}
      \ee
      where $\Phi_\sigma$ is the same as was introduced in
      \cite{9,23},
       $$ \Phi_\sigma (x,y) = P_AP_F \exp (ig \int^x_y
      A_\mu dz_\mu)\times $$ \be \times \exp (g\int^s_0 d\tau
      \sigma_{\mu\nu} F_{\mu\nu}) \label{10.a} \ee and
      $\sigma_{\mu\nu}=\frac{1}{4i}
      (\gamma_\mu\gamma_\nu-\gamma_\nu\gamma_\mu)$, while $K$ and
      ($Dz)_{xy}$ are defined in (\ref{3.a}) and (\ref{4.a}). Note
      that operators $P_A, P_F$ in (\ref{10.a}) preserve the proper
      ordering of matrices $A_\mu$ and $\sigma_{\mu\nu}  F_{\mu\nu}$
      respectively; the explicit examples will be considered below.

      Finally we come to the case of FSR for the valence gluon
      propagating in the background nonabelian field; the detailed
      study of this case is given in chapter 5; here we only quote
      the result (11) for the gluon Green's function in the background
      Feynman gauge
      \be
      G_{\mu\nu} (x,y) =
      \langle x|(D^2_\lambda \delta_{\mu\nu}+
      2ig F_{\mu\nu})^{-1}
      |y\rangle, D_\lambda= \partial_\lambda-ig B\lambda.
      \label{11.a}
      \ee
      Proceeding in the same way as for quarks, one obtains the FSR
      for the gluon Green's function \cite{8}
      \be
      G_{\mu\nu} (x,y) =
      \int^{\infty}_0 ds (Dz)_{xy} e^{-K_0}\Phi_{\mu\nu}(x,y)
      \label{12.a}
      \ee
      where we have defined
      $$
      K_0=\frac14 \int^\infty_0 \left (\frac{dz_\mu}{d\tau}\right)^2
      d\tau;~~ \Phi_{\mu\nu} (x,y) =[P_FP_A\exp (ig \int^x_y A_\lambda
      dz_\lambda)\times
      $$
      \be
      \times \exp (2ig\int^s_0 d\tau F_{\sigma\rho}
      (z(\tau)))]_{\mu\nu}
      \label{13.a}
      \ee
      Now in the same way as it was done above for scalars in
      (\ref{6.a}), one can consider a Green's function for the
      physical transition amplitude from a white state of $q_1, \bar
      q_2$ to another white state consisting of $q_3,\bar q_4$, i.e
      the meson Green's function.
        \be G_{q\bar q} (x,y) =
      \langle G_{q} (x,y) \Gamma G_{\bar q} (x,y)- G_{ q}
              (x,x) \Gamma G_{\bar q} (y,y)\rangle_A \label{14.a} \ee
              and the first term on the r.h.s. of (\ref{14.a}) can be
              reduced to the same form as in (\ref{6.a}) but with the
              Wilson loop containing ordered insertions of the
              operators $\sigma_{\mu\nu} F_{\mu\nu}$
              (cf.(\ref{10.a})).

\section{FSR at nonzero temperature}

This part of lecture is based on papers\cite{17,18}, where  FSR was
first introduced for $T>0$. We discuss first basic formalism at $T>0$
and then gluon and quark  Green's functions.

\subsection{Basic equations}

We start with standard formulas of the background field
formalism \cite{7,8} generalized to the case of nonzero temperature.
We assume that the gluonic field $A_\mu$ can be split into the
background field $B_\mu$ and the quantum field $a_\mu$
\be
A_\mu =B_\mu+a_\mu,
\label{23}
\ee
both satisfying periodic  boundary conditions
$$
B_\mu(z_4,z_i) =B_\mu(z_4+n\beta, z_i),
$$
\be
a_\mu(z_4,z_i) =a_\mu(z_4+n\beta, z_i),
\label{24}
\ee
where $n$ is an integer and $\beta=1/T$.
The partition function can be written as
$$
Z(V,T,\mu=0) =\lan Z(B)\ran_B,
$$
\be
Z(B)=N\int D\phi \exp \left(-\int^\beta_0 d\tau \int d^3 x L_{tot}
(x,\tau)\right ),
\label{25}
\ee
where $\phi$ denotes all set of fields $a_\mu,\Psi, \Psi^+, N$ is
normalization constant, and the sign $\lan~\ran_B$ means some
averaging over (nonperturbative) background fields $B_\mu$, the exact
form of this averaging is not needed for our purposes. Here $L_{tot}$
is the same as in (12)-(15).

Integration over ghost and gluon degrees of freedom in (\ref{25})
yields
$$
Z(B)=N'(\det
W(B))_{red}^{-1/2}[\det(-D_\mu(B)D_\mu(B+a))]_{a=
\frac{\delta}{\delta
J}}\times
$$
\be
\times \left \{1+\sum^\infty_{l=1} \frac{S^l_{int}}{l!} \left( a=
\frac{\delta}{\delta
J}\right)\right\}\exp \left(
-\frac12JW^{-1}J\right)_{J_\mu=D_\mu(B)F_{\mu\nu}(B)}.
\label{26}
\ee
One can consider strong background fields, so that $gB_\mu$ is large
(as compared to $\Lambda^2_{QCD}$), while $\alpha_s= g^2/4\pi$ in
that strong background is small at all distances. Moreover it
was shown that $\alpha_s$ is frozen at large distances
\cite{7,8,25,26}.

                   In this case (\ref{26}) is a perturbative sum in powers of $g^n$,
arising from expansion in $(ga_\mu)^n$.

In what follows we shall discuss the Feynman  graphs for the free
energy $F(T)$, connected to $Z(B)$ via
\be
F(T)=-T\ln \lan Z(B)\ran_B.
\label{27}
\ee
As will be seen, the lowest order graphs  already contain a
nontrivial dynamical mechanism  for the deconfinement transition, and
those will be considered in the next section.

\subsection{The lowest order gluon contribution}

To the lowest order in $ga_\mu$ (and keeping all dependence on
$gB_\mu$ explicitly) one has
\be
Z_0=e^{-F_0(T)/T}=N'\lan \exp (-F_0(B)/T)\ran_B,
\label{28}
\ee
where using (\ref{26}) $F_0(B)$ can be written as
$$
\frac{1}{T} F_0(B) =\frac12\ln \det W-\ln \det (-D^2(B))=
$$
\be
=Sp\left\{ -\frac12
\int^\infty_0 \zeta (t) \frac{dt}{t} e^{-tW} +
\int^\infty_0 \zeta (t) \frac{dt}{t} e^{tD^2(B)}\right\}.
\label{29}
\ee

In (\ref{29}) $Sp$ implies summing over all variables (Lorentz and
color indices and coordinates),  $\zeta(t)= \lim \frac{d}{ds}
\frac{M^{2s}t^s}{\Gamma (s)}|_{s=0}$ is regularizing factor, one
can use also the Pauli-Villars form for $\zeta(t)$.

Graphically, the first term on the r.h.s. of (\ref{29}) and
 is a gluon loop in the background field, the second term is
a ghost loop.

Let us turn now to  the averaging procedure in (\ref{28}). With the
notation $\varphi=-F_0(B)/T$, one can exploit in (\ref{28}) the
cluster expansion \cite{22}
\be
\lan \exp \varphi\ran_B=\exp \left (\sum^\infty_{n=1} \lan \lan
\varphi^n\ran\ran\frac{1}{n!}\right) =\exp \{
\lan
\varphi\ran_B+\frac12[
\lan
\varphi\ran^2_B-
\lan
\varphi^2\ran_B]+O
(\varphi^3)\}.
\label{30}
\ee
To get a closer look at
$\lan
\varphi\ran_B$ we first should discuss thermal propagators of gluon
and ghost in the background field. We start with the thermal ghost
propagator and write the FSR for it \cite{8}
\be
(-D^2)^{-1}_{xy}= \lan x|\int^\infty_0 dte^{tD^2(B)}|y\ran=
\int^\infty_0 dt(Dz)^w_{xy} e^{-K}\hat \Phi(x,y).
\label{31}
\ee
Here $\hat \Phi$ is the parallel transporter in the adjoint
representation along the trajectory of the ghost:
\be
\hat \Phi(x,y) =P\exp (ig \int\hat B_\mu(z) dz_\mu),
\label{32}
\ee
also
$$
K=\frac14\int^t_0d\tau\dot{z}^2_\mu;~~ \dot{z}_\mu=\frac{\partial
z_\mu(\tau)}{\partial\tau}
$$
and $(Dz)^w_{xy}$ is a path integration with boundary conditions
imbedded (this is marked by the subscript $(xy)$) and with all
possible windings in Euclidean temporal direction (marked by the
superscript $w$).

One can write it explicitly as
\be
(Dz)^w_{xy}=
\lim_{N\to \infty}\prod^N_{m=1}
\frac{d^4\zeta(m)}{(4\pi\varepsilon)^2} \sum_{n=0,\pm,...}
\frac{d^4p}{(2\pi)^4}\exp \left[ ip\left( \sum^N_{m=1}
\zeta(m)-(x-y)-n\beta \delta_{\mu 4}\right)\right].  \label{33} \ee

Here, $\zeta(k)=z(k)-z(k-1),~~ N\varepsilon=t$.

One check that in the free case, $ \hat B_\mu=0$ (\ref{31}) reduces
to well-known form of  the free propagator
$$
(-\partial^2)^{-1}_{xy}=\int^\infty_0 dt\exp\left
[-\sum^N_1\frac{\zeta^2(m)}{4\varepsilon}\right ] \prod_m
\overline{d\zeta(m)} \sum_n\frac{d^4p}{(2\pi)^4}\times
$$
$$
 \times \exp \left [ ip \left ( \sum \zeta(m) -(x-y)
 -n\beta\delta_{\mu4}\right)\right] =
 $$
 \be
 =\sum_n\int^\infty_0\exp \left
 [-p^2t-ip(x-z)-ip_4n\beta\right] dt
 \frac{d^4p}{(2\pi)^4}
 \label{34}
 \ee
 with
 $$
 \overline{d\zeta(m)}\equiv \frac{d\zeta(m)}{(4\pi\varepsilon)^2}.
   $$
   Using the Poisson summation formula
   \be
   \frac{1}{2\pi}\sum_{n=0,\pm1, \pm2...} \exp (ip_4 n\beta) =
   \sum_{k=0,\pm1,...} \delta(p_4\beta-2\pi k)
   \label{35}
   \ee
   one finally gets the  standard  form
   \be
   (-\partial^2)^{-1}_{xy}=\sum_{k=0,\pm1,...}
   \int\frac{Td^3p}{(2\pi)^3}\frac{\exp[-ip_i(x-y)_i-i2\pi
   kT(x_4-y_4)]}{ p^2_i+(2\pi kT)^2}.
   \label{36}
   \ee

   Note that, as expected, the propagator (\ref{31}),(\ref{36})
 corresponds to a sum of ghost paths with all possible windings
 around the torus. The momentum integration in (\ref{33}) asserts
 that the sum of all infinitesimal "walks" $\zeta(m)$ should be equal
 to the distance $(x-y)$ modulo $N$ windings in the
 compactified fourth coordinate.

 For the gluon propagator in the background gauge we
 obtain similarly to (\ref{31})
 \be
 (W)^{-1}_{xy}=\int^\infty_0 dt (Dz)^w_{xy} e^{-K}\hat
 \Phi_F(x,y),
 \label{37}
 \ee
 where
 \be   \hat \Phi_F(x,y)= P_F P\exp \left( 2ig \int^t_0 \hat
 F(z(\tau))d\tau\right) \exp \left( ig \int^x_y \hat
 B_\mu dz_\mu\right)
 \label{38}
 \ee
 and the operators $P_F P$ are used to order insertions
 of $\hat F$ on the trajectory of the gluon.

 Now we come back to the first term in (\ref{30}),
 $\lan\varphi\ran_B$, which can be represented with
 the help of (\ref{31})  and (\ref{37}) as
 \be
 \lan \varphi\ran_B=\int\frac{dt}{t} \zeta(t) d^4
 x(Dz)^w_{xx} e^{-K}\left [ \frac12 tr \lan \hat
 \Phi_F(x,x)\ran_B-\lan tr \hat \Phi (x,x)\ran_B\right],
 \label{39}
 \ee
 where the sign $tr$ implies summation over Lorentz and
 color indices.

One can easily show (see \cite{8} for details)
  that (\ref{39}) yields for
 $B_\mu=0$ the usual result for  the free gluon gas:

 \be
 F_0(B=0)=-T\varphi(B=0)=-(N^2_c-1)
 V_3\frac{T^4\pi^2}{45}.
 \label{40}
 \ee

 \subsection{The lowest order quark contribution}

 Integrating over quark fields in (\ref{25}) is done trivially and
 leads to the following additional factor in (\ref{26})
 \be
 \det(m+\hat D(B+a))=\frac12\det (m^2-\hat D^2(B+a)),
 \label{41}
 \ee
 where we have used the symmetry  property of eigenvalues of $\hat
 D$. In the lowest approximation, we omit $a_\mu$ in (\ref{41}) and
 write the free energy contribution $F_0(B)$ similarly to the gluon
 contribution (\ref{28}),(\ref{29}) as
 \be
 \frac{1}{T} F^q_0(B)=-\frac12\ln\det (m^2-\hat D^2(B))=-\frac12
 Sp\int^\infty_0\zeta(t)\frac{dt}{t} e^{-tm^2+t\hat D^2(B)},
 \label{42}
 \ee
 where the sign $Sp$ has the same meaning as in (\ref{29}) and
 \be
 \hat D^2=(D_\mu\gamma_\mu)^2=D^2_\mu(B)+
 gF_{\mu\nu}\sigma_{\mu\nu}
 \equiv D^2+g\sigma F;~~
 \sigma_{\mu\nu} =\frac{1}{4i}
 (\gamma_\mu\gamma_\nu-\gamma_\nu\gamma_\mu).
 \label{43}
 \ee
 Our aim now is to exploit the FSR to represent (\ref{42}) in a form
 of the path integral, as it was done for gluons in (\ref{31}). The
 equivalent form for quarks must implement the antisymmetric boundary
 conditions pertinent to fermions. It is easy to understand that the
 correct form for quark is
 \be
 \frac{1}{T} F^q_0(B)=-\frac12 tr \int^\infty_0\zeta(t) \frac{dt}{t}
 d^4x
 \overline{(Dz)}^w_{xx}e^{-K-tm^2}W_\sigma(C_n),
 \label{44}
  \ee
  where
  $$
  W_\sigma(C_n)=P_FP_A\exp\left (ig \int_{C_n} A_\mu dz_\mu\right)
  \exp g\left (\sigma F\right), $$ $$
 \overline{(Dz)}^w_{xy}=\prod^N_{m=1}\frac{d^4\zeta(m)}{(4\pi
 \varepsilon)^2}\sum_{n=0,\pm1,\pm2,...}
 (-1)^n\frac{d^4p}{(2\pi)^4}
 $$
 \be
 \exp \left [ip\left(\sum^N_{m=1}
 \zeta(m)-(x-y)-n\beta
 \delta_{\mu 4}\right)\right].
 \label{45}
 \ee

 One can easily check that in the case $B_\mu=0$ one is recovering
 the well known expression for the free quark gas
 \be
 F^q_0({\rm free~quark}) =-\frac{7\pi^2}{180} N_cV_3T^4\cdot n_f,
 \label{46}
 \ee
 where $n_f$ is the number of flavors. The derivation of (\ref{46})
 starting from the path-integral form (\ref{44}) is done similarly to
 the gluon case given in the Appendix of \cite{8}.

 The loop $C_n$ in (\ref{44})  corresponds to $n$ windings in the
 fourth direction. Above the deconfinement transition temperature
 $T_c$ one can visualize in (\ref{44}) the appearance of the  factor
 \be
 \Omega=\exp \left [ig \int^\beta_0 B_4(z) dz_4\right].
 \label{47}
 \ee
 (Note the absence of the dash sign in (\ref{47}) as compared
 to(\ref{38}) implying that in (\ref{47}) $B_4$ is taken in
 fundamental representation). For the constant field $B_4$ and
 $B_i=0,i=1,2,3, $ one obtains \be \lan F\ran=-\frac{V_3}{\pi^2}
 tr_c\sum^\infty_{n=1} \frac{\Omega^n+\Omega^{-n}}{n^4} (-1)^{n+1}.
 \label{48}
 \ee
 This result coincides with the obtained in the literature \cite{27}.

 \section{Wilson loop and Field Correlators(FC)}

 Hadron Green's functions in the form of path integrals due to FSR
 are expressed through Wilson loops; moreover all gluonic fields
 enter via these loops, and in case of quarks and gluons additional
 spin factors appear, namely
 \be
 W_F(C)=tr P_A\exp ig \int_{C_0} A_\mu dz_\mu\cdot P_F\exp \int^s_0
 g \sum F(z(\tau)) d\tau
 \label{5.1}
 \ee
 where $\sum F=\sigma_{\mu\nu} F_{\mu\nu}$ for quarks and $ 2i \hat
 F_{\mu\nu}$ for gluons, and the dash sign marks the adjoint
 representation for gluons in contrast to  the fundamental one for
 quarks.

 As it was shown previously in \cite{9,23,28}, the spin factors can
 be treated perturbatively in the effective  Hamiltonian and this
 topic will be discussed in  the last lecture. Here we shall consider
 the simplest form of Wilson loop in the arbitrary representation $D$
 of the $SU(N)$, \be W(C)=tr_DP_A\exp ig \int_{C_0} A_\mu dz_\mu,~~
 A_\mu=A^a_\mu T^a.  \label{5.2} \ee The nonabelian Stokes theorem
 \cite{29} allows to express $W(C_0)$ in terms of field strength
 $F_{\mu\nu},$ instead of the vector potential $A_\mu$, this later
 will enable us to express the vacuum average $\lan W(C_0)\ran$
 through field correlators.

 The easiest way to derive the Stokes theorem is via the
 generalized coordinate gauge \cite{30,31}, namely one introduces the
 following representation for $A_\mu$
 \be
 A_\mu(x)=\int^x_C\frac{\partial z_\rho(s, x)}{\partial x_\mu} dz_\nu
 F_{\nu\rho} (z(s,x))
 \label{5.3}
 \ee
 Here we have introduced an arbitrary contour $C$, which is defined
 by a coordinate $z(s,x)$, starting at $z=x, z(1,x)\equiv x$, while
 for $s=0, z(0,x)=x_0; x_0$ may also lie at infinity.

 The only condition which one should impose on the contour $C(x)$ is
 the "automorfic condition" \cite{31}, i.e.
 \be
 z_\mu(s,x)= z_\mu(s^{\prime\prime}, z(s', x))
 \label{5.4}
 \ee
 which means that choosing on the contour $C(x)$ some point
 $x'\equiv z(s',x)$ to identify it with the initial point of the  new
 contour $C(x')$, one gets $C(x')$ lying entirely on the old contour $C(x)$.

 Inserting now (\ref{5.3}) in (\ref{5.2}) one obtains the nonabelian
 Stokes theorem
 \be
 W(C_0) =tr_D P_S\exp ig \int_{S(C_0)} d\sigma_{\mu\nu} (z)
 F_{\mu\nu}(z)
 \label{5.5}
 \ee
 Note that the ordering $P_S$ in (\ref{5.5}) is fully determined by
 the contour $C(x)$ in (\ref{5.3}) and ordering of points $x$ along
 the contour $C_0$.

 One can show \cite{31} that parallel transporters $\Phi_L(x,y)= P\exp
 ig \int_L A_\mu dz_\mu$ in the generalized coordinate gauge are
 equal to unity whenever contour $L(x,y)$ lies on the contout $C(x)$.
 Therefore one can render the Stokes theorem (\ref{5.5}) to the
 gauge-covariant form
 \be
 W(C_0)=tr_DP_S\exp ig \int_S d\sigma_{\mu\nu} (z) F_{\mu\nu} (z,
 x_0)
 \label{5.6}
 \ee
 where
 \be
 F_{\mu\nu}(z, x_0)\equiv \Phi (x_0, z) F_{\mu\nu} (z) \Phi(z,x_0)
 \label{5.7}
 \ee

The crucial step now is to use the cluster expansion theorem \cite{22}
and to retain only the lowest term, namely
$$
\lan W(C_0)\ran = \lan tr_D P_S\exp ig \int_S d\sigma_{\mu\nu}(z)
F_{\mu\nu} (z,x_0)\ran=
$$
\be
=tr_D\exp \left \{-\frac{g^2}{2}\int_S\int_S d\sigma_{\mu\nu} (z)
d\sigma_{\rho\sigma} (u)\lan\lan F_{\mu\nu} (z,x_0) F_{\rho\sigma}
(u, x_0) \ran\ran +...\right\}.
\label{5.8}
\ee
Here dots imply higher order terms, e.g. $\lan\lan FFF\ran\ran$, etc.
Moreover double brackets mean cumulants, e.g. $\lan\lan AB\ran\ran=
\lan AB\ran-\lan A\ran \lan B\ran$. The angular brackets imply
averaging over all gluonic fields with the standard QCD action, as in
(\ref{1}).

Next step is the standard representation of the bilocal field
correlator in (\ref{5.8}), made in \cite{32}. It goes as follows.
First one realizes, that when $x_0$ is somewhwre in the middle of
surface $S$, then distances $|z-x_0|, |u-x_0|$ are of the order of
$R$ -  "radius" of $S$, while average distance $|z-u|$  will be of
the order of the vacuum correlation length $T_g$.

Hence in the situation when $R\gg T_g$ (and this is typical situation
for hadrons, since as we shall see $R\sim$  size of hadron $\sim 1$
fm, $T_g\sim 0.2$ fm) one can approximate
\be
\lan\lan F_{\mu\nu} (z, x_0) F_{\rho\sigma}(u, x_0)\ran\ran=
\lan F_{\mu\nu}(z)\Phi(z,u) F_{\rho\sigma} (u) \Phi(u,z)\ran+O\left(
\left(\frac{T_g}{R}\right)^2\right)
\label{5.9}
\ee

The lowest correlator (\ref{5.9}) with parallel transporter
connecting points $x$ and $y$ along the straight line has a general
representation in terms of two scalar form-factors $D$ and $D_1$
\cite{32}
$$
\frac{1}{N_c}tr\lan F_{\mu\nu}(x) \Phi(x,y) F_{\lambda\sigma} (y)
\Phi(y,x)\ran=
$$
$$
=(\delta_{\mu\lambda} \delta_{\nu\sigma} -\delta_{\mu\sigma}
\delta_{\nu\lambda})D(x-y) +\frac12 \partial_\mu \{ [
(h_\lambda\delta_{\nu\sigma} - h_\sigma \delta_{\nu\lambda})+
$$
\be
+perm] D_1(x-y)\}, ~~ h_\mu=x_\mu-y_\mu.
\label{5.10}
\ee
It is important here that by definition the term with $D_1$ is a full
derivative. In what follows both functions $D$ and $D_1$ will have a
basic role in all nonperturbative dynamics in QCD, both at zero and
nonzero temperature. To understand why this role is so important one
must compare contribution of the bilocal FC (\ref{5.10}) with all
higher correlators, neglected in (\ref{5.8}).

That comparison can be done at present only indirectly since the
direct determination of higher FC on the lattice is difficult and
still not accurate enough \cite{33}.

Happily there was  another type of lattice measurement recently
\cite{bali1} which has allowed to estimate the contribution  of
higher FC with unprecendented accuracy -better then one percent!

Since the corresponding work \cite{35} is important for understanding
the structure of the QCD vacuum, we shall discuss it is some detail.

Consider static sources in representation $D$ of $SU(3)$ and define
static potential between sources through the Wilson  loop as follows
\be
V_D(R)=-\lim_{T\to \infty} \frac{1}{T} \ln \lan W(C)\ran
\label{5.11}
\ee
where the Wilson loop in the 34 -plane has the form $R\times T$ and
cluster representation
\be
\lan W(C)\ran= Tr_D \exp \sum^\infty_{n=\eta} \int(ig)^n\lan\lan
F(1)... F(n)\ran\ran d\sigma(1) ... d\sigma(n).
\label{5.12}
\ee
Here we have omitted Lorentz indices and denoted
\be
F(n) d\sigma(n)\equiv F(x^{(n)}, x_0) = \Phi (x_0, x^{(n)}) E^a_3
(x^{(n)}) T^a\Phi(x^{(n)}, x_0) d\sigma_{34}(x)
\label{5.13}
\ee

The $SU(3)$ representations $D=3,8,6,15a, 10,27, 24, 15s$
are characterized by $3^2-1=8$ hermitian generators $T^a$ which
satisfy the commutation relations $[T^a, T^b] = i f^{abc} T^c$.
One of the main characteristics of the representation
is an eigenvalue of quadratic Casimir operator ${\cal C}^{(2)}_D $,
which is defined according to ${\cal C}^{(2)}_D = T^a T^a = C_D\cdot \hat1$.
Following the notations from \cite{bali1} we
introduce the Casimir ratio $d_D = C_D / C_F$, where
the fundamental Casimir $C_F = (N_c^2-1)/2N_c$ equals to
$4/3$ for $SU(3)$. The invariant trace is given by $\T_D \hat1 =1$.

Since a simple algebra of the rank $k$ has exactly $k$ primitive
Casimir--Racah operators \cite{cr}
of order $m_1,..,m_k$, it is possible to
express
those of higher order in terms of the primitive ones.
In the case of $SU(3)$ the primitive Casimir operators
are given by
\be
{\cal C}^{(2)}_D = {\delta}_{ab} T^a T^b \;\; ; \;\;
{\cal C}^{(3)}_D = {d}_{abc} T^a T^b T^c
\label{eq67}
\ee
while the higher rank Casimir operators are defined as follows
\be
{\cal C}^{(r)}_D = d^{(r)}_{(i_1 .. i_r)} T^{i_1} .. T^{i_r}
\ee
where the totally symmetric tensor $d^{(r)}_{(i_1 .. i_r)}$
on the $SU(N_c)$ is expressed in terms of $\delta_{ik}$ and
$d_{ijk}$ (see, for example,
\cite{hysp}).

The potential (\ref{5.11}) with the definition (\ref{5.12}) admits the
following decomposition
 \be
 V_D(R)= d_D V^{(2)}(R) + d^2_D V^{(4)}(R)+...,
 \label{eq3}
 \ee
where the part denoted by dots contains terms, proportional to
the higher powers
of the quadratic Casimir as well as to higher Casimirs.

The fundamental static potential contains perturbative
Coulomb part,  confining linear and constant
terms
\be
V_D(R) = \sigma_D R - v_D - \frac{e_D}{R}
\label{pppo}
\ee
 The  Coulomb part is now known up to two loops \cite{psr}
and is proportional to $C_D$.
The "Casimir scaling hypothesis"
\cite{go} declares, that
the confinement potential is also proportional to the
first power of the quadratic Casimir $C_D$, i.e.
all terms in the r.h.s. of (\ref{eq3}) are much smaller than
the first one. In particular, for the string tensions
one should get $\sigma_D/\sigma_F = d_D$.

This scaling law is in perfect agreement with the results
found in \cite{bali1,35}. Earlier lattice calculations
of static potential between sources in higher representations
\cite{go}
are in general agreement
with \cite{bali1}.

To see, why this result (to be more
precise -- why the impressive {\it accuracy} of the "Casimir scaling"
behaviour) is nontrivial,
let us examine the colour structure of a few lowest
averages in the expansion (\ref{5.12}).

The first nontrivial Gaussian cumulant in
(\ref{5.12}) is expressed through
$C_D$ and representation--independent averages as
\be
\T_D\lll F(1) F(2)\rrr =
\frac{C_D}{N_c^2 -1}\> \lll F^a(1)
F^a(2)\rrr =
\frac{d_D}{2N_c}\> \lll F^a(1)
F^a(2)\rrr,
\label{eq4}
\ee
so Gaussian approximation satisfies "Casimir scaling law"
exactly. It is worth being mentioned, that this fact does not depend
on the actual profile of the potential.
It could happen, that the linear potential observed
in \cite{bali1} is just some kind of
intermediate distance characteristics and changes the profile
at larger $R$ (as it actually should happen
in the quenched case for the
representation of zero triality due to the screening of
the static sources by dynamical gluons from the vacuum, or,
in other words, due to gluelumps formation).
The coordinate dependence
of the potential
is not directly related to the Casimir scaling,
 and can be analized
at the distances which are small enough
to be affected by the screening effects.

Having made these general statements, let us come back to
our analysis of the contributions to the potential
from different field correlators.
We turn to the quartic correlator and
write below several possible colour structures
for it. We introduce the following abbreviation
$$
\lll F^{[4]} \rrr = \lll F^a(1) T^a  F^b(2) T^b F^c(3) T^c
F^d(4) T^d \rrr
= T^a T^b T^c T^d {\lll F^{[4]} \rrr}^{abcd}
$$
where the Lorentz indices and coordinate dependence
are omitted for simplicity of notation.
One then gets, with some work in the last case
the following possible structures
$$
\begin{array}{ll}
{\lll F^{[4]}\rrr}^{abcd} \sim
{\delta}_{ab}\delta_{cd}&
\lll F^{[4]} \rrr \sim C_D^2 \cdot \hat1 \\
{\lll F^{[4]} \rrr}^{abcd} \sim
{\delta}_{ac}\delta_{bd} &
\lll F^{[4]} \rrr \sim \left(C_D^2 -\frac{1}{2} N_c C_D\right)\cdot
\hat1 \\
{\lll F^{[4]} \rrr}^{abcd} \sim
{f}_{ade} f_{cbe}&
\lll F^{[4]} \rrr \sim -\frac14 N_c^2 C_D \cdot \hat1 \\
{\lll F^{[4]} \rrr}^{abcd} \sim
{f}_{ace} f_{bde} &
\lll F^{[4]} \rrr = 0 \\
{\lll F^{[4]} \rrr}^{abcd} \sim
{f}_{apm} f_{bpn} f_{dem} f_{cen}\;\;\;\;\; &
\lll F^{[4]} \rrr \sim \frac94 \left(
C_D^2 + \frac12 C_D
\right) \cdot \hat1 \\
\end{array}
$$

The operator ${\cal C}^{(3)}_D$ enters together with ${\cal C}^{(2)}_D$
at higher orders.
Notice, that the terms, proportional to the
square of $C_D$ appear in both the $\delta \delta$ parts
(the first and the second strings)
and higher order interaction parts (the last string).
Mnemonically the $C_D$ -- proportional components
arise from the diagrams where the noncompensated
colour flows inside
while the $C_D^2$ -- components describe the interaction
of two white objects. It is seen, that the Casimir
scaling does not mean "quasifree gluons",
instead it means roughly speaking "quasifree white multipoles"
(see discussion at the end of the paper).

Let us analyse the data from \cite{bali1} quantitatively.
We have already mentioned, that the Coulomb potential
between static sources
is proportional to $C_D$ up to the second loop (and possibly
to all orders, this point calls for further study) and hence
we expect contributions proportional to
$C^2_D\sim d^2_D$ to the constant
  and linear terms, i.e. we rewrite (\ref{pppo})
as follows
  \be
  V_D(R)= d_D V^{(2)}(R)+ d^2_D (v_D^{(4)} +  \sigma_D^{(4)} R).
  \label{eq5}
  \ee
and all higher contributions are omitted.
  Here
   $ v_D^{(4)}, \sigma_D^{(4)}$
   measure the $d_D^2$--contri\-bu\-tion of the
  cumulants higher than Gaussian to the constant term and string
  tension respectively.
The results of the fitting of the data from \cite{bali1} with
(\ref{eq5}) for some representations are shown in the Table 1.

\begin{table}
\caption{ {The Casimir--scaling and Casimir--violating string
tensions
and shifts.
Based on the data from G.Bali, hep-lat/9908021. [34]}}
\bigskip
\bigskip

\begin{tabular}{|c|c|c|c|c|c|c|}
\hline
& & & & & & \\
$D$ & $ {\sigma}_D^{(4)}\cdot 10^{4}$ & $\Delta
{\sigma}_D^{(4)}\cdot 10^{4}$
& $v_D^{(4)}\cdot 10^{4}$ & $\Delta v_D^{(4)}\cdot 10^{4}$ & $
\left| \sigma_D^{(4)}/ \sigma_D^{(2)}\right| $& $ \chi^2 / N $
\\
& & & & & &\\
\hline
& & & & & &\\
8 &  -3.486  &   1.2 & -2.513 & 2.8 & 0.004 & 19.22 / 43
\\
& & & & & &\\
\hline
& & & & & &\\
  6 & -6.428 &  1.2 & 0.950  &  2.6 &0.007 & 25.76 / 42
\\
& & & & & &\\
\hline
& & & & & &\\
15a   &-5.244  & 0.55   &-0.5611  &1.1 &0.003   &39.06 /
42 \\
& & & & & &\\
\hline
   &  &   &  & &  &  \\
10 &-4.925 &  0.50 & 0.2489 & 1.0 &0.003 & 22.05 / 41\\
& & & & & &\\
\hline
\end{tabular}
\label{tab2}
\end{table}
All numbers in the Table 1 are dimensionless
and given in lattice
units.
The author of \cite{bali1} used anisotropic lattice with the spatial
unit $a_s^{-1} = 2.4 GeV $.

Several comments are in order. First of all it is seen
that the Casimir scaling behaviour holds with very good
accuracy, better than 1\% in all cases in the table 1
with the reasonable $\chi^2$. It should be stressed, that
any possible systematic errors which could be present
in the procedure used in \cite{bali1} must either obey
the Casimir scaling too or be very tiny, otherwise
it would be unnatural to have the matching with such
high precision.\footnote{Since we are mostly interested in the relative
quantities, their actual magnitude in the physical units
is of no prime importance for us. This is another reason
why we do not discuss
possible systematical errors of \cite{bali1} and finite
volume effects. }
Nevertheless, the terms violating the scaling are also
clearly seen. While the value of the constant term $v_D^{(4)}$ is
found to be compatible with zero within the error bars,
it is not the case for $\sigma_D^{(4)}$. We have not found
any sharp dependence of $\sigma_D^{(4)}$ on the representation $D$,
which confirms the validity of the expansion (\ref{eq5})
and shows, that the omitted higher terms do not have
significant effect in this case. Notice the negative
sign of the string tension
correction. In euclidean metric it
trivially follows from the fact, that the fourth order
contribution is proportional to $(ig)^4 >0$ while the
Gaussian term is multiplied by $(ig)^2 <0$ for real $g$.

The analysis given above allows to draw important conclusions about
the QCD vacuum:

1) The contribution of higher FC to the static potential is less than
1\% at all measured distances $0.1 fm\leq R\leq 1.3 fm$. (Another
possible scenario is that all higher FC are fine-tuned to cancel
exactly all terms proportional to $C^n_D, n>1$; but this fine-tuning
seems rather artificial).

2) The lowest correlators $D$ and $D_1$ should describe dynamics of
the Wilson loop (i.e. the dynamics of nonperturbative QCD with
 accuracy better than 1\%. This fact can be checked in calculations
 of physical amplitudes, e.g. we shall give exmples of our
 predictions in the last lecture.

 3) All modes of the QCD vacuum which are based on classical
 solutions are strongly limited by the observed Casimir scaling,
 since a classical lump (e.g. an instanton or  a dyon) generate a big
 Quartic (and higher) FC.

 E.g. the density of instanton gas (in the SU(2) case) in the QCD
 vacuum has an upper bound of $\frac{N}{V}=0.2 fm^{-4}$ (and even
 smaller if one considers smaller distances). This should be compared
 with the normal density of instanton gas of $1 fm^{-4}$ providing
 the standard gluonic condensate.

 The picture of the QCD vacuum which emerges from the Casimir scaling
 is the picture of stochastic ensemble of small white dipoles, where
 only any two adjacent points are correlated, while all the rest are
 statistically independent. This picture is ideal for description in
 terms of lowest field correlators, $D(x)$ and $D_1(x)$ at all
 temperatures, and we shall indeed show how easily one can describe
 dynamics of different precesses in QCD.

 4) As a by-product the results of [34,35] give a strong upper bound
 on the so-called Luescher-Weissz term in the potential \cite{40},
 $V_{LW} =-\frac{\pi}{12R}$,
 since the latter violates Casimir  scaling. This means that the QCD
 string is either not the Nambu-Goto string, used in \cite{40}, or
 else  one should use  properly the finite lattice size for vibrating
 strings -- this is an interesting problem for the future.

 \section{Field correlators at zero and nonzero temperature}

 Since $D(x)$ and $D_1(x)$ are basic elements of the nonperturbative
 QCD dynamics, it is important to find them as functions of distance
 $x$ and use later on as an input.

 The only standard way  to get information on $D$, $D_1$ before was
 the lattice simulation method, and lattice measurements of $D,D_1$
 have been done both at $T=0$ \cite{41}  at $T>0$ \cite{42}.

 At the end of these lectures we shall describe other methods of
 calculating field correlators and in particular $D$ and $D_1$, but
 now we shall describe in detail results of the Pisa group.

 For $T=0$ these data are presented in Fig.1, and can be parametrized
 by the form
 \be
 D(x)=Ae^{\mu x}+ \frac{a e^{-\mu_1x}}{x^4}
 \label{5.14}
 \ee
 \be
 D_1(x)=Be^{\mu' x}+ \frac{b e^{-\mu'_1x}}{x^4}
 \label{5.15}
 \ee

 The last terms in (\ref{5.14}), (\ref{5.15}) proportional to
 $x^{-4}$ are clearly of perturbative origin,
       moderated at large distances by screening factors.
       The most important information is contained in the first terms
       in (\ref{5.14}), (\ref{5.15}) proportional to exponents. It
       is striking that numerically $\mu=\mu'\simeq 1$ GeV  for the
       unquenched case.

       At the end of the lectures we shall explain the appearance of
       exponents in (\ref{5.14}). (\ref{5.15}) and the mass values
       around  1 GeV -- this will be connected to the masses of the
       so-called gluelumps -- bound states of gluons with a static
       adjoint source. These masses have been calculated recently
       both on the lattice \cite{43} and analytically \cite{44}.

       Physically appearance of  large masses - of the order of 1 GeV
       leads to very important consequences:

       i) First, the gluon correlation length $T_g\equiv 1/\mu\approx
       0.2 fm$ is small as compared to the generic hadronic size and
       implies  that  stochastic picture of small independent white
       dipoles which was discussed in previous section.

       ii) As a consequence of small $T_g$ one can develop the
       potential picture of hadrons (see \cite{23,28} for more
       discussion), with quarks and valence gluons interacting via a
       string instantaneously without retardation,
       while nonpotential corrections are proportional to some powers
       of $(T_g/R_h)$, where $R_h$ is  the typical hadron size.

       iii) Finally, smallness of $T_g$ puts serious limitation on
       the use of some standard methods in QCD. E.g. the application
       of QCD sum rules strictly speaking implies large values of
       $T_g$, namely $T_g\gg R_h$, otherwise the notion of gluon
       condensate and its derivatives loses its  meaning.

       The same is true for the Voloshin-Leutwyler method, based on
       the constant condensate contribution to the heavy quarkonia
       dynamics. One can estimate that the inequality $R_{Q\bar
       Q}\leq T_g$ holds true  for toponium and hardly for
       bottomonium.

       Knowing $D$ and $D_1$ one can calculate all the dynamics of
       heavy quarkonia (and also of other hadrons -- in the next
       sections).

       In particular the string tension is defined by the
       nonperturbative part of $D(x)$ \cite{32}
       \be
       \sigma=\frac12\int d^2 xD(x)
       \label{5.16}
       \ee
       Note that $D_1$ does not contribute, being the full
       derivative, and the perturbative part of $D(x)$ is cancelled
       by the corresponding perturbative parts of triple and higher
       correlators \cite{45}. The whole static potential, as expressed
       through $D$ and $D_1$ has the form
       \be
       V(R) = 2R
       \int^R_0 d\lambda\int^\infty _0 d\nu D(\lambda, \nu)+
       \int^R_0\lambda d\lambda\int^\infty _0 d\nu[-2 D(\lambda,
       \nu)+ D_1(\lambda, \nu)]
       \label{5.17}
       \ee
Here $D(\lambda, \nu) = D(\sqrt{\lambda^2+\nu^2})$ because of
Euclidean $O(4)$ invariance. Note that the first term on the r.h.s.
of (\ref{5.17}) yields at large $R$ the term $\sigma R$, while the
second term adds there a constant. The perturbative contribution to
$V(R)$ can be included in $D_1$, or considered separately, in both
cases one obtains the form
\be V(R)= \sigma R-v
-4/3\frac{\alpha_s(R)}{R}
 \label{5.18} \ee observed many times on
the lattice and recently in \cite{bali1}.

We shall comment in what follows on the NP dependence of
$\alpha_s(R)$, but now we turn to spin-dependent  corrections to the
potential, calculated in \cite{28,46}, i.e. spin-orbit potentials
$V_1, V_2$, tensor force $V_3$ and spin-spin potential $V_4$ (we
exploit the standard notations of Eichten and Feinberg \cite{47}).
All these potentials are expressed through $D$ and $D_1$ only and in
this way the dynamics  of heavy  quarkonia is fully determined.

The corresponding calculations  have been done repeatedly and
successfully during last 10 years in \cite{48}-\cite{51}. In the
Tables 2 and 3, given are the known levels of charmonium and
bottomonium calculated using the lattice data on $D, D_1$ from
in a standard way. One can see a very good description of all
levels, which far exceeds in quality the newly developed  NRQCD
methods.

In the Table 2 listed are bottomonium level differences, which have
more physical significance than the absolute values of levels, since
 the heavy quark mass is not known and used as an input parameter.
 Calculations in \cite{51} have been done both in relativistic and
 nonrelativistic kinematics, and one can estimate therefore the
 accuracy of 1/M expansion usually done in the framework of NRQCD.
 One can see that corrections can be very significant, especially for
 $P$-wave levels, where the matrix element
 of the $1/r^3$ operator is increased up
 to 20\% in relativistic calculations
 \cite{51}.

 Similar calculations have been done  for charmonium in \cite{50},
 again with very good results. An interesting comparison of these and
 similar results is contained in \cite{52},  where the
 Voloshin-Leutwyler method was used, and shown to be applicable (if
 any) only to ground state of bottomonium.

\begin{table}[h]
\caption{ Bottomonium level differences (MeV) for the
Schr\"{o}dinger and Salpeter equations from \cite{51}
}
\bigskip
\bigskip

\begin{tabular}{|c|c|c|c|c|c|}
\hline
Masses differences&
 \multicolumn{2}{c|}{Set I, $\alpha_{eff}=0.3545$}&
 \multicolumn{2}{c|}{ Set II, $\alpha_{eff}=0.36$}&  Exp.val\\
  &\multicolumn{2}{c|}{$m=4.737$ GeV} &
   \multicolumn{2}{c|}{$m=4.81 $GeV } &(MeV)\\
   &\multicolumn{2}{c|}{$\sigma=0.20 $ GeV $^2 $}
& \multicolumn{2}{c|}{$\sigma=0.18 $ GeV $^2$} &\\ \hline
&Rel.&Nonrel.&Rel.&Nonrel.&\\\hline
$M(2S)-M(1S)$&
554.34 &551.97 &556.55 &550.03 &562.9$\pm$0.5\\
$M(3S)-M(2S)$&
 350.43 &354.78 &335.62 &338.49 &332.0$\pm$0.8\\
$M(4S)-M(3S)(^a)$&
 285.93 &291.83 &270.63 &275.30 &224.7$\pm$ 4.0\\
$M(1P)-M(1S)$&
 458.04 &439.66 &473.49 &450.15 &439.8$\pm$0.9\\
$M(2P)-M(1P)$&
 359.67 &366.75 &342.55 &348.70 &359.8$\pm$1.2\\
$M(2S)-M(1P)$&
 96.31 &112.31 &83.07 &99.88 &123.1$\pm$1.0\\
$M(3S)-M(2P)$&
 87.06& 100.34 &72.82 &89.67 &95.3$\pm$1.0\\
 \hline
\end{tabular}
 \label{tab3}
  \end{table}

 $(^a)$ The $4S$ level lies above the $b\bar B$ threshold and the
 corresponding threshold correction was not taken into account, which
 explains the discrepancy.\\
\newpage

\begin{table}[h]
\caption{ The spin-averaged mass level differences in
charmonium $^{a)}$ (in MeV)~~~~~~~~~~~~~~~~~~~~~~~~~~
 }
 \bigskip
\bigskip

\begin{tabular}{|c|c|c|c|}
\hline
&&Set$C$&Set $D$\\
&&$M=1.48$ GeV,& $m=1.40$ GeV,\\
& Experimental
& $\sigma=0.18$ GeV$^2$,
& $\sigma=0.183$ GeV$^2$,\\
&values$^{b)}$&
$\alpha_s(\mu_0)=0.365,$&
$\alpha_s(\mu_0)=0.312,$\\
&(MeV)&
$\mu_0=0.909 $ GeV,&
$\mu_0=1.357 $ GeV,\\
&&$\tilde \alpha_V=0.42$&
$\tilde \alpha_V=0.39$\\\hline
$M(2S)-M(1S)$&
595.39 $\pm$ 1.91&  591.30& 586.09\\
  $M(1P)-M(1S)$& 457.92 $\pm$ 1.0& 460.68& 445.93\\
  $M(2P)-M(1P)$& 137.47$\pm $1.77& 130.62& 140.15\\\hline
  \end{tabular} \label{tab4}
 \end{table}

 $^{a)}$ all matrix elements defining spin effects were
 calculated for spinless Salpeter equation.\\

 $^{b)}$ $M(nL)$ means the spin-averaged mass.

 \section{The Field Correlator Method for light quarks}

 Let us start with the meson Green's function and use for that FSR
 for  both quark and antiquark.

\be
G_M(x,y)=\langle tr \Gamma^{(f)}(m-\hat D)\int^\infty_0 ds \int^\infty_0
d\bar s e^{-K-\bar K}(Dz)_{xy}(D\bar z)_{xy}\Gamma^{(i)}(\bar
m-\hat{\bar D}) W_F\rangle
\label{87}
\ee
Here the bar sign refers to the antiquark, and
\be
W_F=P_A P_F \exp(ig\oint dz_\mu A_\mu) \exp(g
\int^s_0
d\tau\sigma^{(1)}_{\mu\nu}F_{\mu\nu})
\exp (-g\int^{\bar s}_0\sigma^{(2)}_{\mu\nu}F_{\mu\nu} d\bar\tau)
\label{88}
\ee
In (\ref{87}) enter integrations over proper times $s, \bar s$ and
$\tau, \bar\tau$, which also play the role of ordering parameter
along the trajectory, $z_\mu=z_\mu(\tau),~~\bar z_\mu=\bar z_\mu(\bar
\tau)$.

It is convenient to go over to the actual time $t\equiv z_4$ of the
quark (or antiquark), defining the new quantity $\mu(t)$, which will
play very important role in what follows
\be
2\mu(t) =
\frac{dt}{d\tau},~~ t\equiv z_4(\tau)
\label{89}
\ee
For each quark (or antiquark and gluon) one can rewrite the path
integral (\ref{12.a}), (\ref{87}) as follows
\be \int^\infty_0
ds (D^4z)_{xy}...  = const \int D\mu(t) (D^3z)_{xy}...
\label{90}
  \ee where
$(D^3z)_{xy}$ has the same form as in (\ref{4.a}) but with all
4-vectors replaced by 3-vectors, and the path integral $D\mu(t)$ is
supplied with the proper integration measure, which is  derived from
the free motion Lagrangian.

In general $\mu(t)$ can be a strongly oscillating function of $t$ due
to the Zitterbewegung. In what follows we shall use the steepest
descent method for the evaluation of the integral over $D\mu(t)$,
with the extremal $\mu_0(t)$ playing the role of effective or
constituent quark mass. We shall see that in all cases, where spin
terms can be considered as a small perturbation, i.e. for majority of
mesons, $\mu_0$ is positive and rather large even for vanishing quark
current masses $m,\bar m$, and the role of Zitterbewegung is small
(less than 10\% from the comparison to the light-cone Hamiltonian
eigenvalues, see \cite{53,54} for details).

Now the kinetic terms can be rewritten using (\ref{89}) as
\be
K+\bar K=\int^T_0 dt\{ \frac{m^2}
{2\mu(t)} +\frac{\mu(t)}{2} [(\dot
z_i(t))^2+1]+
\frac{\bar m^2}{2\bar \mu(t)} +\frac{\bar \mu(t)}{2} [(\dot{\bar
z_i}(t))^2+1]\}
\label{91}
\ee
where $T=x_4-y_4$. In the spin-dependent factors the corresponding
changes are
\be
\int^s_0 d\tau \sigma_{\mu\nu} F_{\mu\nu}=
\int^T_0 \frac {dt}{2\mu(t)}\sigma_{\mu\nu} F_{\mu\nu}(z(t)).
\label{92}
\ee
In what follows in this section  we shall systematically do
perturbation expansion of the spin terms,
which contribute to the total mass corrections of the order of
10-15\% for lowest mesons and much smaller for high excited states.
This perturbative approach fails however for pions (and kaons) where
chiral degrees of freedom should  be taken into account. In this case
other equation should be considered \cite{55,56}.  Therefore as the starting
approximation we shall use the Green's functions of mesons made of
spinless quarks, which amounts to neglect in (\ref{87}),(\ref{88})
terms $(m-\hat D), (\bar m-\hat {\bar D})$ and
$\sigma_{\mu\nu}F_{\mu\nu}$.  As a result one has \be G^{(0)}_M(x,y)=
const \int D\mu(t) D\bar \mu(t) (D^3z)_{xy} (D^3\bar z)_{xy}
e^{-K-\bar K}\langle W\rangle .  \label{93} \ee

 Our next approximation is the neglect of perturbative exchanges in
 $\langle W\rangle $,
 (they will be restored in the final expression for Hamiltonian)
 which yields for large Wilson loops, $R,T\gg T_g$,
 \be
 \langle W\rangle _B=const~ exp(-\sigma S_{min})
 \label{94}
 \ee
 where $S_{min}$ is the minimal area inside the given trajectories of
 quark and antiquark,
 \be
 S_{min} =\int^T_0 dt\int^1_0 d\beta \sqrt{detg},~~
 g_{ab}=\partial_aw_\mu\partial_bw^{\mu},~a,~b=t,~\beta.
 \label{95}
 \ee

 The Nambu-Goto form of $S_{min}$ cannot be quantized due to the
 square root and one has to use the auxiliary field approach \cite{57}
 with functions $\nu(\beta,t)$ and $\eta(\beta,t)$ to get rid of the
 square root, as it is usual in string theories. As the result the
 total Euclidean action becomes \cite{58}
 $$
 A=K+\bar K+\sigma S_{min}=
 \int^T_0 dt\int^1_0 d\beta
 \{\frac12(\frac{m^2}{\mu(t)}+\frac{\bar m^2}{\bar
 \mu(t)})+\frac{\mu_+(t)}{2}\dot R^2+
 $$
 \be
 +\frac{\tilde \mu(t)}{2}\dot r^2+
 \frac{\nu}{2}[\dot w^2+(\frac\sigma\nu)^2r^2-2\eta (\dot
 wr)+\eta^2r^2]\}.
 \label{96}
 \ee

 Here $\mu_+=\mu+\bar
\mu,~~\tilde\mu=\frac{\mu\bar\mu}{\mu+\bar\mu},~~
 R_i=\frac{\mu z_i+\bar\mu\bar z_i}{\mu+\bar\mu},~~
 r_i=z_i-\bar z_i$. Note, that integrations over $\mu,\nu$ and $\eta$
 effectively amount to the replacement by their extremum values
 \cite{58}.

 Performing Gaussian integrations over $R_{\mu}$ and $\eta$ one
 arrives in the standard way at the Hamiltonian (we take $m=\bar m$
 for simplicity)
$$
H=\frac{p^2_r+m^2}{\mu(\tau)}+
\mu(\tau)+\frac{\hat L^2/r^2}
{\mu+2\int^1_0(\beta-\frac{1}{2})^2\nu(\beta) d\beta}+
$$
\be
+\frac{\sigma^2
r^2}{2}\int^1_0\frac{d\beta}{\nu(\beta)}+
\int^1_0\frac{\nu(\beta)}{2}d\beta,
\label{17.a}
\ee
where $p^2_r=(\vep\ver)^2/r^2$, and $L$ is the angular momentum,
$\hat L=(\ver\times \vep)$.

The physical meaning of the terms $\mu(t)$ and $\nu(\beta)$ can be
understood when one finds their extremal values. E.g. when $\sigma=0$
and $L=0$, one finds from (\ref{17.a})
\be
H_0=2\sqrt{\vep^2+m^2},~~\mu_0=\sqrt{\vep^2+m^2}
\label{97}
\ee
so that $\mu_0$ is the energy of the quark. Similarly in the limiting
case $L\to \infty$ the extremum over $\nu(\beta)$ yields:
\be
\nu_0(\beta)=\frac{\sigma r}{\sqrt{1-4y^2(\beta-\frac{1}{2})^2}},
~~H^2_0=2\pi\sigma\sqrt{L(L+1)}
\label{98}
\ee
so that $\nu_0$ is the energy density along the string with the
$\beta$ playing the role of the coordinate along the string.

Let us start with the $L=0$ case. Taking extremum in $\nu(\beta)$ one
has
\be
H_1=\frac{p^2_r+m^2}{\mu(t)}
+\mu(t)+\sigma r.
\label{23.a}
\ee
Here $\mu(t)$ is to be found also from the extremum amd is therefore
an operator in Hamiltonian formalism. Taking extremum in $\mu(t)$ one
obtains
\be
H_{2}=2\sqrt{p_r^2+m^2}+\sigma r.
\label{24.a}
\ee
The Hamiltonian (\ref{24.a}) is what one traditionally exploits in the
relativistic quark model (RQM) \cite{59} (apart from color
Coulomb and spin-dependent terms to be discussed below).  The RQM was
an essential step in our understanding of hadronic spectra.
At the same time the usual input of  RQM contains too
many parameters and the model  was introduced rather ad hoc.  Another
deficiency of this model at this point is twofold:

i) one usually takes $m$ in (\ref{97}),(\ref{24.a}) to be constituent
quark mass of the order of 100-200 MeV, which is introduced as an
input.  Instead we have in (\ref{17.a}),(\ref{98}),(\ref{24.a}) the
current quark mass renormalized at the reasonable scale of 1 GeV.
Hence it is almost zero for light quarks;

ii) the form (\ref{24.a}) is used in RQM for any $L$, and to this end
one writes in (\ref{24.a}) $p^2_r \to \vep^2$.

However the Regge slope for both (\ref{23.a}) and (\ref{24.a}) is
$1/8\sigma$ instead of the string slope $\frac{1}{2\pi\sigma}$,
which occurs for the total Hamiltonian (\ref{17.a}), since the RQM
Hamiltonian (\ref{24.a}) does not take into account string rotation.

Still for $L=0$ the form (\ref{24.a}) is a good starting
approximation, and it is rewarding that our systematic approach makes
here contact with RQM. Sometimes it is convenient to use instead of
(\ref{24.a}) a more tractable form (\ref{23.a}) where $\mu(t)= \mu_0$
and $\mu_0$ is to be found from the extremum of the eigenvalue of
Hamiltonian $H_1$:  \be
H_1\psi=M_1\psi,~~M_1=M_1(\mu_0),~~\frac{\partial M_1}{\partial
\mu_0}=0.
\label{99}
\ee

Equation (\ref{23.a}), (\ref{24.a}) can be reduced to the Airy
equation with the eigenvalues written as (for $m=0$) \be M_n^{(1)}
=4\mu_0(n),~~ \mu_0(n)=\sqrt{\sigma}(\frac{a(n)}{3})^{3/4}
\label{100}
 \ee where
$a(n)$ is the corresponding zero of the Airy function.
The first comprehensive  study of the
equation (\ref{99}) with fixed $\mu(t)\equiv m$ was done in \cite{60},
where also $a(n)$ are quoted.

It is easy to apply the Hamiltonian (\ref{17.a}) to the case of light,
\cite{23,60,61} heavy \cite{48}-\cite{51} and heavy-light mesons
{\cite{24,62}.  We refer here to \cite{61} where these
applications and also hybrid calculations  have been successfully
done.  Again, the most important point of the described formalism is
the natural occurence of the constituent mass of quark and gluon,
which is calculated through the string tension (and $\alpha_s$) in
the unique way, as shown in Eq.(\ref{99}).

Since this point is very important from both phenomenological and
physical point of view, we shall discuss here in more detail the
case of baryon magnetic moment (defined through quark constituent
mass) and spectrum of glueballs  (where constituent mass
of gluons enter in a decisive way).

For baryons the Hamiltonian is derived from the Green's function of a
baryon in the same way as it was done above for mesons.

We refer the reader to \cite{63}-\cite{65} for details of derivation.
The result is
 \be H=\sum^3_{k=1} \left (
\frac{m^2_k}{2\mu_k}+\frac{\mu_k}{2}\right)+\left(-
\frac{\partial^2}{\partial\vexi^2}-
\frac{\partial^2}{\partial\veta^2}\right) +\sigma \sum^3_{k=1}
|\ver^{(k)}|.
\label{101}
\ee

Here          $m_k$ are current quark masses defined at the scale of
1 GeV, $\mu_k$ - the corresponding constituent masses (energies) to
be calculated in what follows, $\ver^{(k)}$ is the  vector from
position of the $k$-th quark to the string junction. The coordinates
$\vexi, \veta$ are Jacobi combinations defined through $\ver^{(k)}$
in a standard way.

The eigenvalues of (\ref{101}) $M_n(\mu)$ can be found in a simple way
with  good accuracy using the  the hyper-spherical explansion method
(menthod of the $K$ - harmonics) \cite{66}. Assuming equal masses
$M_k=m$ and hence equal constituent masses $\mu_k=\mu$, one obtains
equation for the definition of $\mu$
\be
\frac{\partial M_n(\mu)}{\partial\mu}
\left|_{\mu=\mu^{(0)}}=0\right..
\label{102}
\ee
>From that we find $\mu\equiv \mu_u$ for baryon consisting of 3 light
quarks
\be
\mu_u=2\sqrt{\frac{2\sigma}{\pi}}\left[ \frac{2}{3\cdot 5^{1/3}}
\left(1+\frac{2}{3\sqrt{5}}\right)\right]^{3/4} =0.957\sqrt{\sigma}.
\label{103}
\ee

Choosing $\sigma =$0.15 GeV$^2$ (as it accepted for baryons
phenomenologically \cite{67} and recently found microscopically
\cite{68}) one has $\mu_u=0.37$ GeV. For the strange quarks,
$m_s=0.175$ GeV, one instead has $\mu_s=0.46$ GeV.

Now comes the most dramatic moment, since $\mu_u,\mu_s$ are directly
connected to the baryon magnetic moment. Namely, similarly to the
color-magnetic term, also interaction with usual magnetic field has
the form
 \be
 \delta
 A=\sum^3_{k=1}\int^{s_k}_0 d\tau_k e_k\vesig^{(k)}{\bf B}=
 \sum^3_{k=1}\int^T_0\frac{e_k{\vesig}^{(k)}{\bf
 B}}{2\mu_k} dt,
 \label{13.b}
 \ee
 where $e_k$ is the electric charge of the quark,
 ${\vesig}^{(k)}$ is the corresponding spin operator, and the
 definition (\ref{89}) of the constituent mass was used.

 Introducing the $z$-component of the magnetic moment operator
 \be
\hat\mu_z
= \sum^3_{k=1}\frac{e_k{\sigma}^{(k)}_z}{2\mu_k},
 \label{0.14}
 \ee
 one can write the BMM as matrix elements
 \be
 {\mu}_B\equiv \lan \Psi_B|
 {\hat\mu}_z|\Psi_B\ran
 \label{14.b}
 \ee
 where $\Psi_B$ is the eigenfunction of (\ref{101}), and $\mu_k$ is
 taken at the stationary point,
 given by (\ref{102}).

For the baryon wave function we shall take here the simplest
approximation, namely
\be
\Psi_B=
\Psi^{symm}(r) \psi^{symm}(\sigma, f) \psi^a(color),
\label{15.b}
\ee
where $\psi(\sigma, f)$ is the spin-flavour part of the wave
function.

           Results of calculations of baryon magnetic moments
           \cite{65} are given in Table 4 in nuclear magnetons.

\begin{table}[h]
\caption{
 Magnetic moments of baryons (in nuclear magnetons)
  computed using Eqs.(\ref{11}),(\ref{0.14}),(\ref{14}) in
  comparison with experimental data from PDG \cite{24}}
\bigskip
\bigskip

  \begin{tabular}{|l|l|l|l|l|l|l|l|l|l|l|}\hline
Baryon&$p$&$n$&
$\Lambda$&$\Sigma^-$&$\Sigma^0$&$\Sigma^+$&
$\Xi^-$&$\Xi^0$&$\Omega^-$&$ \Delta^{++}$\\\hline
Present work&2.54 &-1.69 &-0.69& -0.90 &0.80
&2.48&-0.63&-1.49&-2.04& 436\\\hline
Experiment&2.79 &-1.91&-0.61&-1.16&&2.46&-0.65&-1.25&-2.02& 4.52\\
\hline \end{tabular}
 \end{table}

Note that all our predicted values of $\mu^{(B)}$ are calculated only
through the string tension and this is the first
calculation of this sort, which tells that the string picture of
baryons (we call it the QCD string model) is very good starting
approximation -- agreement of the calculations with experiment is on
average better than of any other approach.

One more example in this section will illustrate the universality of
the approach, where the notion of constituent gluon mass enters and
this quantity is explicitly calculated.

The  example is the glueball spectrum, calculated in this
approach in \cite{69,70}.

Since gluons have no current masses, $m_k=0$, the hamiltonian for
angular momentum $L=0$ has the form (without spin and $L$
corrections)
\be
H_0=\frac{\vep^2}{\mu_0} +\mu_0+\sigma_{adj} r
\label{107}
\ee
Here $\sigma_{adj}=\frac94\sigma_{fund},~~ \sigma_{fund}=0.18 $
GeV$^2$. The eigenvalues are easily obtained from (\ref{100}),
\be
M_0(n)=4\mu_0(n),~~
\mu_0(n)=\sqrt{\sigma_{adj}}\left(\frac{a(n)}{3}\right)^{3/4},
\label{108}
\ee
and $a(L=0,n=0)=2.338,~~
a(L=1,n=0)=3.36,~~
a(L=0,n=1)=4.088.$

To this Hamiltonian $H_0$ one should add spin-dependent terms $H_s$
(totally expressed through $\alpha_s$ and  $D,D_1
 see \cite{61} and refs. therein)$ and orbital
correction $H_L$ derived in \cite{58}. Referring the reader to
\cite{69,70} for  details here we only give the Table of galueball
masses in comparison with two sets of lattice calculations (for the
same value of $\sigma_f=0.238 $ GeV$^2$, chosen on the lattice).

\begin{table}[h]
\caption{
Comparison of predicted glueball masses with lattice data (for
$\sigma_f=0.238$ GeV$^2  $ )       }
\bigskip
\bigskip

 \begin{tabular}{|l|l|l|l|} \hline
$J^{PC}$& $M$ (GeV)&  \multicolumn{2}{c|}{Lattice data}\\\cline{3-4}
 &This work& Ref. [1]&Ref. [3]\\\hline

$0^{++}$&1.58&1.73$\pm$0.13&1.74$\pm$0.05\\
$0^{++*}$&2.71&2.67$\pm$0.31&3.14$\pm$0.10\\
$2^{++}$&2.59&2.40$\pm$0.15&2.47$\pm$0.08\\
$2^{++*}$&3.73&3.29$\pm$0.16&3.21$\pm$0.35\\
$0^{-+}$&2.56&2.59$\pm$0.17&2.37$\pm$0.27\\
$0^{-+*}$&3.77& 3.64$\pm$0.24&          \\
$2^{-+}$&3.03&3.1$\pm$0.18&3.37$\pm$0.31\\
$2^{-+*}$&4.15& 3.89$\pm$0.23&          \\
$3^{++}$&3.58&3.69$\pm$0.22&4.3$\pm$0.34\\
$1^{--}$&3.49& 3.85$\pm$0.24&          \\
$2^{--}$&3.71& 3.93$\pm$0.23&          \\
$3^{--}$&4.03& 4.13$\pm$0.29&          \\\hline
  \end{tabular}
 \end{table}

One can notice a remarkable agreement between analytic and lattice
data. Note also that both calculations have  essentially the same
input: the string tension $\sigma_f$, while spin splittings are also
defined by the value of the strong coupling $\alpha_s$. It means that
the Field Correlator Method is also successful in describing gluon
bound system and the notion of valence gluon and its constituent mass
(energy) is fully supported by independent calculations.

Let us turn now to another gluon system, which has important
theoretical rather than  experimental meaning. This is the gluelump
system calculated both on the lattice \cite{43} and in the framework
of the present method in \cite{44}. There is a remarkable agreement
between the states defined in the same way (through space-like
links).

The most important feature of gluelumps is that the lowest masses
with electric quantum numbers $1^{--}$ correspond exactly to the
masses observed in the exponents by the Pisa group \cite{41}. And
indeed calculations in \cite{44} for the electric gluelump (which is
described by electric correlator $\lan E_i \Phi E_k\ran $) yield for
the mass $\mu\approx 1$ GeV in agreement with \cite{41}. Note however
that magnetic gluelumps with quantum numbers  $1^{+-}$, probably
could not been observed in \cite{41}, since quantum numbers were not
specifically projected in \cite{41}, and therefore presumably only
lowest mass could be observed.

Analytic calculation of gluelumps, i.e. of field correlators with
given  given quantum numbers, signifies  a new important period in
the development of the Method as whole: the functions $D(x) $ and
$D_1(x)$ which have been before introduced as input from lattice
data, will be calculated analytically within the same method, making
finally the whole theory complete and selfconsistent. The first
attempts in this direction have been already done in \cite{6}, where
$\mu=1/Tg$ was calculated in terms of string tension.

\section{Dynamics of the temperature phase transition in QCD}

Calculations of the free energy of gluons and quarks, which have been
done in section 4, will serve  us in this section to compute
the phase transition temperature and to define the physical picture
of the phase transition - in the first approximation of $1/N_c$
expansion - which will be a very realistic picture. .

We shall consider two possible phases - the deconfined one at
$T>T_c$ and the confined phase at $T<T_c$.
 The choice of the phase  from physical point of view is
dictated by the minimal value of the free energy (or the maximal
value of pressure).
The most important ingredient in the QCD free energy (which is absent
in QED) is the nonperturbative energy density, connected to the
gluonic condensate
\be
\varepsilon= \frac{\beta(\alpha_s)}{16\alpha_s} \lan F^a_{\mu\nu}
F^a_{\mu\nu}\ran \simeq -\frac{11}{3} N_c\frac{\alpha_s}{16\pi} \lan
E^aE^a+H^a H^a\ran.
\label{111}
\ee
Thus one write the free energy in the low phase (confined) as
\be
\frac{1}{V_3} F_{low} =\varepsilon-
T\sum_k\frac{(2m_kT)^{3/2}}{8\pi^{3/2}} e^{-m_k/T}-\frac{\pi^2}{30}
T^4+O(1/N_c)
\label{112}
\ee
where $m_k$ are massive particles (not pions), pions are given by the
third term on the r.h.s. of (\ref{112}).

For deconfined phase one should specify the condensate term
(\ref{111}). Since confinement is electric, deconfinement would
logically imply disappearence of the electric part of the condensate
$\lan E^a E^a\ran$ in (\ref{111}). However the NP part of $D_1(x)$
does not confine and therefore in $\lan E^a E^a\ran =D^E(0)
+D_1(0)$ the part $D_1^E(0)$ could survive. However lattice
calculations of the Pisa group \cite{42} imply, that both
$D^E(0)$ and $D_1^E(0)$ vahish near $T=T_c$, while the magnetic
part $D^B(0)+D^B_1(0)$ stay intact and contribute around one half
of the total condensate, as it is for zero temperature.

Hence one can write the free energy for the high phase as
\be
\frac{1}{V_3} F_{high} =\frac12 \varepsilon - (N^2_c-1)
\frac{T^4\pi^2}{45} -\frac{7\pi^2}{180} N_cT^4 n_f
\label{113}
\ee
where $n_f$ is the number of flavours, and the second and the
third term refer to gluons and quarks computed respectively in
(\ref{40}) and (\ref{5.8}).

Comparing (\ref{112}) and (\ref{113})
one can see that $F_{high}$ starts higher than $F_{low}$ at $T=0$
(note that $\varepsilon<0$) but drops faster and at some $T=T_c$ both
$F$ are equal, which means that the phase transition takes place and
nature chooses that phase which ensures lower free energy (higher
pressure) (this is a consequence of the 2nd thermodynamic law).

This value of $T_c$ is \cite{17,18}.
\be
T_c=\left[ \frac{\frac{11}{6} N_c\frac{\alpha_s\lan F^\ran}{32\pi}}
{\frac{\pi^2}{45} (N^2_c-1)+\frac{7\pi^2}{180} N_c n_f}\right]^{1/4}.
\label{114}
\ee
Note that at large $N_c$ the critical temperature is $O(N_c^0) $,
which is expected, but no other model predicts this correct
dependence.

It is instructive to compare the values of $T_c$ from (\ref{114}) and
from recent lattice calculations \cite{71}.

\begin{table}[h]
\caption{
 Values of $T_c$ for $N_c=3$ and different $n_f$ compared to
lattice data \cite{71}}
\bigskip
\bigskip

 \begin{tabular}{|l|l|l|}\hline
 $SU(3)$&   $T_c(m_\rho=0.77)$ (MeV)& Eq.(\ref{114}) MeV\\ \hline
 $n_f=0~^{*)}$ &270& 220-240\\
 $n_f=2$ &170& 150\\
 $n_f=4$ &131& 134\\  \hline
 \end{tabular}
 \end{table}

 $ ^{*)} $ For $n_f=0$ the quenched value of the gluonic condensate
 was taken 3-4 times larger than the standard (nonquenched) value
 $\frac{\alpha_s}{\pi}\lan F^2\ran =0.012 $ GeV$^4$.
 This in agreement with arguments of SVZ\cite{2} and lattice
 calculations of Di Giacomo et al \cite{42}.

 Several comments are in order.  First, values of $T_c$ are
 surprisingly close to the lattice calculations which supports our
 picture of phase transition. Secondly, magnetic condensate is as
 strong as for $T=0$ (and becomes stronger with grouth of $T$). This
 leads to the spacial correlations above $T_c$, calculated in
 \cite{20} and measured on the lattice.

 Finally, one should not confuse the above physically correct picture
 of phase transition with a popular model, where all system is
 immerged in the huge bag with the standard bag pressure (which is
 order of magnitude smaller than gluonic condensate \cite{2}). This
 latter picture leads to many inconsistencies and cannot be
 considered as a serious suggestion.

 Recent development of our picture demonstrating a good description
 of nonperturbative dynamics of quark gluon plasma was done for
 $SU(2)$ in \cite{21} and quite recently for $SU(3)$ in \cite{72}.
 Further work incorporating both electric field in Polyakov lines and
 magnetic field modifying gluon and quark free energy is now under
 investigation.

 \section{Acnowledgements}

 The author is grateful to the Organizing Commitee of the
 International School for a kind hospitality and good creative
 atmosphere during the whole period.

\end{document}